\documentclass{llncs}
\usepackage{wrapfig}
\usepackage{llncsdoc}
\usepackage{epsfig,psfrag}
\usepackage{latexsym,amssymb,amsmath}
\usepackage{longtable}
\usepackage{tikz,pgf}
\usepackage{subfig}
\usepackage{wrapfig}
\usepackage{rotating}

\usepackage{graphicx}
\usepackage{epstopdf}
\usepackage{color}

\usepackage{amsmath,amsfonts}
\newcommand{\bqn}{\begin{eqnarray}}
\newcommand{\eqn}{\end{eqnarray}}
\newcommand{\bq}{\begin{eqnarray*}}
\newcommand{\eq}{\end{eqnarray*}}

\usepackage{url}
\usepackage{hyperref}
\hypersetup{colorlinks,%
citecolor=black,%
filecolor=black,%
linkcolor=red,%
urlcolor=blue,%
pdftex}

\begin{document}

\title{Rapid Acceleration of the Permutation Test via Slow Random Walks in  the Permutation Group}

\author{Moo K. Chung$^1$,  Yixian Wang$^1$, Shih-Gu Huang$^1$,
Ilwoo Lyu$^3$}

\institute{
University of Wisconsin, Madison, USA\\
Vanderbilt University, Nashville, USA\\
\vspace{0.3cm}
}

\maketitle


\begin{abstract}
The permutation test is an often used test procedure in brain imaging. Unfortunately, generating every possible permutation for large-scale brain image datasets such as HCP and ADNI with hundreds images is not practical. Many previous attempts at speeding up the permutation test rely on various approximation strategies such as estimating the tail distribution with known parametric distributions. In this study, we show how to rapidly accelerate the permutation test without any type of approximate strategies by exploiting the underlying algebraic structure of the permutation group. The method is applied to large number of MRIs  in two applications: (1) localizing the male and female differences and (2) localizing the regions of high genetic heritability in the sulcal and gyral pattern of the human cortical brain.

\end{abstract}

\section{Introduction}
The permutation test  is perhaps the most widely used nonparametric test procedure in sciences \cite{chung.2017.IPMI,thompson.2001,zalesky.2010,nichols.2002}. It is known as the exact test in statistics 
since the distribution of the test statistic under the null hypothesis can be exactly computed if we can calculate all possible values of the test statistic under every possible permutation. 
Unfortunately, generating every possible permutation  for whole images is still extremely time consuming even for modest sample size.

When the total number of permutations is too large, various resampling techniques have been proposed to speed up the computation in the past. In the resampling methods, only a small fraction of possible permutations are generated and the statistical significance is computed {\em approximately.} This approximate permutation test is the most widely used version of the permutation test. In most of brain imaging studies, 5000-1000000 permutations are often used, which puts the total number of generated permutations usually less than a fraction of all possible permutations. In \cite{zalesky.2010}, 5000 permutations are out of possible ${27 \choose 12}=17383860$ permutations (2.9\%) were used. In \cite{thompson.2001}, 1 million permutations out of ${40 \choose 20}$ possible permutations (0.07\%) were generated using a super computer.  
This is an approximate method and a care should be taken to guarantee the convergence but in most studies about 1\% of total permutations are used, mainly due to the computational bottleneck of generating permutations \cite{thompson.2001,zalesky.2010}.

To remedy the computational bottleneck, the tail regions of the distributions are often estimated using the extreme value theory, which deals with modeling extreme events and rare occurrences \cite{smith.1989,embrechts.1999}.  One main tool in the extreme value theory is the use of  generalized Pareto distribution in approximating the tail distributions at high thresholds., which requires far smaller number of permutations. Unfortunately, without a prior information or model fit, it is difficult to even guess the shape of tails accurately. Recently, an exact combinatorial approach with quadratic run time was proposed in \cite{chung.2017.IPMI}, but the method is limited to a KS-distance on monotone features not applicable to more general test statistics.  

In this paper, we propose a novel {\em slow random walk} strategy for permutations that rapidly accelerates the permutation test without any approximation. Unlike the traditional permutation test that takes up to few days in a single computer, our method takes less than an hour in a laptop. Our main contributions of this paper are as follows. 1)  New slow random walk approach to rapidly accelerate the permutation test. Theoretically, our approach is few hundred times faster, which is equivalent to having hundreds more computers. 
2) The proposed framework is applied to a large sample study in differentiating brain images of males and females. As far as we are aware, this study is perhaps the largest sample size study where the permutation test was performed. 3) The method is further used in showing how to accurately compute the twin correlations in ACE genetic model on permutations on twin labels.

\section{Preliminary}
The usual permutation test setting in applications is a two-sample comparison \cite{nichols.2002,thompson.2001}. 
Consider  two ordered sets ${\bf x} = ( x_1, x_2, \cdots, x_m)$ and ${\bf y}  = (  y_1, y_2, \cdots, y_n )$. The distance between ${\bf x}$ and ${\bf y}$ is measured by test statistic $f({\bf x},{\bf y})$ such as $t$-statistics or correlations. Under the null hypothesis of equivalence of ${\bf x}$ and ${\bf y}$, elements in ${\bf x}$ and ${\bf y}$ are permutable.

Consider the combined ordered set ${\bf z} = ( x_1, \cdots, x_m, y_1, \cdots, y_n )$ and its all possible permutations $\mathbb{S}_{m+n}$. Since there is an isomorphism between ${\bf z}$ and integer set $\{1, 2, \cdots, m+n \}$, we will interchangeably use them when appropriate. $\mathbb{S}_{m+n}$ is a symmetric group of order $m+n$ with $(m+n)!$ possible permutations \cite{kondor.2007}. For $ \tau \in \mathbb{S}_{m+n}$, it is often denoted as
$$\tau = 
\left(
\begin{array}{cccccc}
  x_1 &  \cdots & x_m  & y_1 & \cdots &  y_n\\
\tau(x_1) & \cdots & \tau(x_{m}) & \tau(y_1) &  \cdots & \tau(y_n)   \\ 
\end{array}
\right).$$

For instance, consider a permutation of $\{1, 2, 3, 4\}$ given by 
$\tau(1) = 4, \tau(2)=2, \tau(3)=1, \tau(4)=3$. Since there are two cycles in the permutation, $\tau$ can be written in the {\em cyclic form} as $\tau =  [2][1,4,3]$ indicating 2 is a cycle of length 1 ($2 \to 2)$ while 1, 3, 4 are a cycle of length 3 ($1 \to 4 \to 3 \to 1$)  \cite{hungerford.1980}. If the meaning is clear, a cycle of length 1 is simply ignored and the permutation is written as $\tau=[1,4,3]$. If another permutation is given by $\pi(1)=1,\pi(2)=4,\pi(3)=3,\pi(4)=2$, the sequential application of $\pi$ to $\tau$  is written as
$$\pi \cdot \tau = [1][3][2,4] \cdot [2][1, 4, 3] = [2,4]\cdot[1,4,3]=[1, 2, 4, 3].$$

Let us split the permutation $\tau({\bf z})$  into two groups with $m$ and $n$ elements
$$ \tau({\bf x}) = (\tau(x_1), \cdots, \tau(x_m) ), \quad \tau({\bf y}) = ( \tau(y_1), \cdots, \tau(y_n) ).$$
Let $f(\tau({\bf x}),\tau({\bf x}))$  be a test statistic, which measures the distance between the permuted groups. The {\em exact} $p$-value for testing an one sided hypothesis is then given by the fraction
\bqn p\mbox{-value} =\frac{1}{(m+n)!} \sum_{\tau \in \mathbb{S}_{m+n}  } \mathcal{I} \Big(  f(\tau({\bf x}),\tau({\bf y}))  > f ({\bf x},{\bf y}) \Big),  \label{eq:pvalue} \eqn
where $\mathcal{I}$ is an indicator variable taking value 1 if the argument is true and 0 otherwise. In various brain imaging applications, computing statistic $f$ for each permutation has been the main computational bottleneck \cite{nichols.2002,thompson.2001,chung.2017.IPMI}.

If the test statistic $f$ is a symmetric function in each argument such that 
$$
f({\bf x},{\bf y}) = f(\phi( {\bf x}),\psi({\bf y})),$$
where $\phi \in \mathbb{S}_{m}$ and $\psi \in \mathbb{S}_{n}$, then due to the multiplicity,  we only need to consider ${m+n \choose m} = {m+n \choose n}$ number of permutations in the denominator of (\ref{eq:pvalue}), which reduces the number of possible permutations substantially. Still ${m+n \choose m}$ is an extremely large number and most computing systems including MATLAB cannot compute them exactly if the sample size is larger than 100 in each group. The total number of permutations in permuting two vectors of size $m$ each is given asymptotically by Stirling's formula ${2m \choose m} \sim 4^m/\sqrt{2\pi m}$   \cite{chung.2017.IPMI}. The number of permutations {\em exponentially} increases as the sample size increases, and thus it is impractical to generate every possible permutation. In practice, up to hundreds of thousands of random permutations are generated using the uniform distribution randomly on  $\mathbb{S}_{m+n}$ with probability $1/{m+n \choose m}$.

\section{Methods}
{\em Core idea.} In the standard permutation test, half of $m$ and $n$ objects are randomly chosen and mixed in each permutation in average. Since there is no rule in how each object is chosen, test statistic $f$ has to be computed using all the objects in each permutation. If we sequentially construct permutations  in such a way that we only pick one object per group from the previous permutation, it is possible to achieve the constant run time $O(const.)$ in computing test statistic $f$. This slowdown of the mixing proportion in the permutations  is the key to rapidly accelerating the permutation test.

\subsection{Random walks on the permutation group}

Consider permutation $\pi_{ij}$ that exchanges $i$-th and $j$-th elements between ${\bf x}$ and ${\bf y}$ and keeps all others fixed such that
\bq \pi_{ij}({\bf x}) &=& (x_1, x_2, \cdots, x_{i-1}, y_j, x_{i+1}, \cdots, x_m), \label{eq:x} \\
 \pi_{ij}({\bf y}) &=& (y_1, y_2, \cdots, y_{j-1}, x_i, y_{j+1}, \cdots, y_n).  \label{eq:y} \eq
Such a permutation is called the {\em transposition} or  {\em walk}. Walks on the permutation group is related to card shuffling problems and it is a special case of  walk in symmetric groups \cite{aldous.1983}. 
The walk between elements {\em within} ${\bf x}$ or ${\bf y}$ is also allowed but will not affect the computation for symmetric test functions. Consider every possible sequence of walks applied to ${\bf x}$ and ${\bf  y}$. If such sequence of walks covers every possible element in $\mathbb{S}_{m+n}$, we can perform the permutation test by sequentially transposing two elements at a time. 

\begin{theorem} Any permutation in $\mathbb{S}_{m+n}$ can be reachable by  a sequence of walks. 
\end{theorem}
{\em Proof.}  Here we provide the sketch of proof \cite{hungerford.1980}. Let $l=m+n$. Suppose $\tau \in \mathbb{S}_{l}$. For $x \in \{1, \cdots, l \}$, consider cycle 
$$C_x = [x, \tau(x), \tau^2(x), \cdots, \tau^j (x) ]$$ 
with $\tau^{j+1}(x) =x$ and $\tau^d(x) \neq x$ for $d \leq j$. Since we are dealing with the finite number of elements, such $j$ always exists. 
If $\tau^c(x) = \tau^d(x)$ for some $c \leq d \leq j$, we have $\tau^{d-c}(x) = x$, thus all elements in the cycle $C_x$ are distinct.
If $C_x$ covers all the elements in $\{1, \cdots, l\}$ we proved the statement. If there is an element, say $y \in \{1, \cdots, l\}$, that is not covered by $C_x$, we construct a new cycle $C_y$. Cycles $C_x$ and $C_y$ must be disjoint. If not, we have $\tau^i(x) = \tau^j(y)$ and $y=\tau^{i-j}(x)$, which is in contradiction.
Hence $\tau = C_x \cdot C_y$. If $C_x \cdot C_y$ does not cover 
$\{1, \cdots, l\}$, we repeat the process until we exhaust all the elements in $\{1, \cdots, l\}$. Hence any permutation can be decomposed as a product of {\em disjoint} cycles. Then algebraic derivation can further show that cycle $C_x$ can be decomposed as a product of 2-cycles 
$$C_x = [x, \tau^j(x)]\cdot[x, \tau^{j-1}(x)] \cdots [x ,\tau^2(x)]\cdot[x , \tau(x)]. $$
2-cycle is simply a walk. Hence we proved $\tau$ is a sequence of walks. 
$\qed$

\subsection{Statistics over random walks}
Now we know it is possible to obtain any permutation by a sequence of walks. Instead of performing uniform random sampling in $\mathbb{S}_{m+n}$, we will perform a sequence of random walks and compute the test statistic at each walk.  

Consider walks in the two sample setting. We will determine how test statistic changes over each walk. Suppose $i$-th and $j$-th entries are transposed between ${\bf x}$ and ${\bf y}$ such that
\bq \pi_{ij}({\bf x}) &=& (x_1, x_2, \cdots, x_{i-1}, y_j, x_{i+1}, \cdots, x_m),\\
 \pi_{ij}({\bf y}) &=& (y_1, y_2, \cdots, y_{j-1}, x_i, y_{j+1}, \cdots, y_n).\eq
Over random walk $\pi_{ij}$, the statistic changes from $f({\bf x}, {\bf y})$ to $f( \pi_{ij} ({\bf x}), \pi_{ij}({\bf y}))$. Instead of computing $f( \pi_{ij} ({\bf x}), \pi_{ij}({\bf y}))$ directly, we will compute it from $f({\bf x}, {\bf y})$ incrementally in {\em constant} run time by updating  the value of $f({\bf x}, {\bf y})$. 

\begin{theorem}
\label{theorem:algebraic}
 If $f$ is an algebraic function that only involves addition, subtraction, multiplication, division, integer exponents, there exists a function $g$ such that 
\bqn f(\pi_{ij}({\bf x}), \pi_{ij}({\bf y})) = g (f ({\bf x}, {\bf y}), x_i, y_i), \label{eq:increment} \eqn
where the computational complexity of $g$ is constant. 
\end{theorem}
The lengthy proof involves explicitly constructing iterative formula for each algebraic operations so it will not be shown here. However, the correctness of the statement will be obvious after studying two examples in the next sections. Often used statistics such as the two-sample $t$-statistic and $F$-statistic are algebraic functions. If we take computation involving {\em fractional exponents} as constant run time as well, then a much wider class of statistics such as correlations can all have iterative formulation (\ref{eq:increment}) with constant run time. 

{\em Computational complexity.} Suppose we are computing two-sample $t$-statistic with $m$ and $n$ samples directly. This requires computing the sample means, which is $O(m)$ and $O(n)$ algebraic operations each. Then need to compute the sample variances and pool them together, which requires $O(3m+2)$ and $O(3n+2)$ operations each. Combining the numerator and denominator in $t$-statistic takes $O(16)$ operations. Thus, it takes total $O(4(m+n)+20)$ operations to compute the $t$-statistic at each permutation. In comparison, we show that it is possible to achieve the constant run time $O(const.)$ for each permutation using the proposed random walk.

\subsection{Two sample $t$-statistic}

Let the sample mean and variance of ${\bf x}$ and ${\bf y}$ be \bq \mu ({\bf x}) &=& \frac{1}{m}\sum_{k=1}^m x_k, \quad \sigma^2({\bf x}) = \frac{1}{m-1} \sum_{k=1}^m (x_k-\mu({\bf x}))^2\\
 \mu ({\bf y}) &=& \frac{1}{n}\sum_{k=1}^n y_k,  \quad \sigma^2({\bf y}) = \frac{1}{n-1} \sum_{k=1}^n (y_k-\mu({\bf y}))^2.
 \eq
Suppose $i$-th and $j$-th entries are transposed between ${\bf x}$ and ${\bf y}$. The means after transposition can be written as 
$$m\mu (\pi_{ij}({\bf x}))= m\mu({\bf x})+ y_j-x_i, 
\quad 
n\mu (\pi_{ij}({\bf y}))= n\mu({\bf y})+ x_i-y_j.$$
The variances after transposition can be written as
\bq (m-1)\sigma^2 (\pi_{ij}({\bf x})) &=& (m-1)\sigma^2({\bf x})+ \Big(\big(m\mu ({\bf x})\big)^2-  \big(m\mu (\pi_{ij}({\bf x}))\big)^2\Big)/m + y_j^2-x_i^2,\\
(n-1)\sigma^2(\pi_{ij}({\bf y})) &=& (n-1) \sigma^2({\bf y})+ \Big(\big(n\mu ({\bf y})\big)^2-  \big(n\mu (\pi_{ij}({\bf y}))\big)^2\Big)/n+ x_i^2-y_j^2.
\eq
The iterative update of means and variances  will compound numerical errors if we keep dividing by $m$ and $n$ over sequential random walks. Thus, we simply update 
$$\nu ({\bf x}) = m \mu ({\bf x}), \nu ({\bf y}) = n \mu ({\bf y}), 
\omega^2 ({\bf x}) = (m-1) \sigma^2 ({\bf x}), \omega^2 ({\bf y}) = (n-1) \sigma^2 ({\bf y}).$$
The two-sample $t$-statistic after transposition is then computed as
$$T(\pi_{ij}({\bf x}),\pi_{ij}({\bf y}))=\Big( \frac{\nu (\pi_{ij}({\bf x}))}{m}-\frac{\nu (\pi_{ij}({\bf y}))}{n} \Big) \Big/\sqrt{ \frac{
\omega^2 (\pi_{ij}({\bf x})) + \omega^2 (\pi_{ij}({\bf y})) } { m+n -2}  \Big(\frac{1}{m} + \frac{1}{n} \Big)}.$$

{\em Computational complexity.} Computing $\nu (\pi_{ij}({\bf x}))$ and $\nu (\pi_{ij}({\bf y}))$ requires $O(2)$ operation each. Computing $\omega^2 (\pi_{ij}({\bf x}))$ and $\omega^2 (\pi_{ij}({\bf y}))$ needs  $O(9)$ operations each. Then combining the numerators and denominators to construct $t$-statistic needs $O(13)$ operations. Thus the total computational complexity of updating $t$-statistic over a walk is $O(35)$. \\

\begin{wrapfigure}{r}{0.43\textwidth}
\vspace{-2cm}
\centering
\includegraphics[width=1\linewidth]{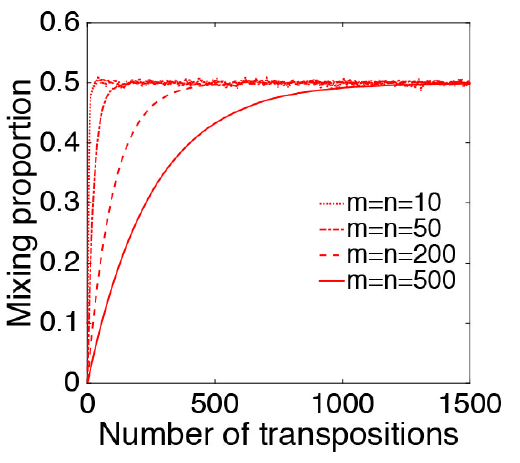}
 \caption{\small The estimated mixing proportion based on the average of 1000 sequences of random transpositions.
 }\label{fig:mixing}
 \vspace{-0.5cm}
\end{wrapfigure}

{\em Mixing proportion.} Given $m=n$ elements in each group, the standard permutation test mixes half of elements in one group to the other. Thus, the mixing proportion is 0.5 in average. On the other, the transposition method mixes one element at a time, so the mixing is slow but it rapidly catches up. For instance, for sample sizes $m=n=200$, about 500 transpositions are enough to mix all the elements in the two groups uniformly (Figure \ref{fig:mixing}).  For far smaller sample sizes, which most brain imaging studies belong, few hundreds transpositions are enough.

\subsection{Twin correlation estimation}

It is often necessary to compute the twin Pearson correlation within MZ- or DZ-twins in the ACE genetic model  as baselines \cite{chung.2017.IPMI,falconer.1995}. Consider twin data 
$${\bf x}=(x_1, x_2, \cdots, x_n), \; {\bf y}=(y_1, y_2, \cdots, y_n),$$
where
$(x_i,y_i)$ is the $i$-th paired twin. The twin correlation is denoted as $\rho({\bf x}, {\bf y})$.
Since there is no preference in the order of twins, we can transpose the $i$-th twin pairs such
\bq \pi_{i}({\bf x}) &=& (x_1, x_2, \cdots, x_{i-1}, y_i, x_{i+1}, \cdots, x_n),\\
 \pi_i({\bf y}) &=& (y_1, y_2, \cdots, y_{i-1}, x_i, y_{i+1}, \cdots, y_n),\eq
and obtain another twin correlation $\rho(\pi_i({\bf x}), \pi_i({\bf y}))$. There are $2^n$ possible correlation combinations and this causes in ambiguity in the ACE model. To remedy the problem, we will randomly transpose the twins in pairwise fashion and compute twin correlations. Then take the average of up $2^n$ correlation values  as the robust estimate for twin correlation. For large $n$, this is astronomically large and causes a serious computational bottleneck. Thus we will use the proposed random walk approach.

The sample means $\mu$ after transpositions are updated as
$$n\mu (\pi_i({\bf x}))= n \mu({\bf x})+ y_i-x_i, 
\quad 
n \mu (\pi_i({\bf y}))= n \mu({\bf y})+ x_i-y_i.$$
The sample variances   $\sigma^2$  after transpositions are written as
\bq (n-1) \sigma^2 (\pi_i({\bf x})) &=& (n-1) \sigma^2(\textbf x)+  \Big(\big(n\mu ({\bf x})\big)^2-\big(n\mu (\pi_i({\bf x}))\big)^2\Big)/n+ y_i^2-x_i^2,\\
(n-1) \sigma^2(\pi_i(\textbf y)) &=& (n-1) \sigma^2(\textbf y)+  \Big(\big(n\mu ({\bf y})\big)^2-\big(n\mu (\pi_i({\bf y}))\big)^2\Big)/n+ x_i^2-y_i^2.\eq
The sample covariance $\sigma$ between  $\pi_{i}({\bf x})$ and $ \pi_{i}({\bf y})$ can be written as
\bq
n(n-1) \sigma(\pi_i({\bf x}), \pi_i ({\bf y}))= n(n-1) \sigma(\textbf x, \textbf y)+  (x_i-y_i)^2+(x_i-y_i) \big(n\mu({\bf y}) - n\mu({\bf x}) \big).
\eq
The iterative update of means and variances will compound numerical errors if we keep dividing by $n$ in each walk. Thus, we simply update 
\bq \nu ({\bf x}) &=& n \mu ({\bf x}), \; \; \; \nu ({\bf y}) = n \mu ({\bf y}), \\
\omega^2 ({\bf x}) &=& (n-1) \sigma^2 ({\bf x}), \; \; \; \omega^2 ({\bf y}) = (n-1) \sigma^2 ({\bf y}),\; \; \; \omega({\bf x}, {\bf y}) = n(n-1)\sigma({\bf x}, {\bf y}).
 \eq
Then the correlation after transpositions is computed iteratively as
\bq
\rho(\pi_i({\bf x}), \pi_i({\bf y}))&=&\frac{\sigma(\pi_i({\bf x}), \pi_i ({\bf y})) }{\sqrt{\sigma^2(\pi_i({\bf x})) \sigma^2(\pi_i({\bf y}))}}\\
&=&\frac{\omega(\pi_i({\bf x}), \pi_i({\bf y}))}{n\sqrt{\omega^2 (\pi_i({\bf x}))\omega^2 (\pi_i({\bf y}))}}
\eq

{\em Computational complexity.} 
Computing $\nu (\pi_{i}({\bf x}))$ and $\nu (\pi_{i}({\bf y}))$ requires $O(2)$ operations each. Computing $\omega^2 (\pi_{i}({\bf x}))$ and $\omega^2 (\pi_{i}({\bf y}))$ needs  $O(9)$ operations each. Computing covariance $\omega(\pi_i({\bf x}), \pi_i({\bf y}))$ needs $O(7)$ operations. Then combining them into  correlation needs $O(4)$ operations. Thus the total time complexity is $O(33)$.

\subsection{Multiple comparisons}
So far we have shown how test statistics change over random walks. Here, we show how $p$-values and multiple comparison corrected $p$-values change over random walks.  Suppose ${\bf x}(q)$ and ${\bf y}(q)$ are functional data on position $q$ in some brain region $\mathcal{M}$. Given statistic map $f(q) = f({\bf x}(q), {\bf y}(q))$, the hypotheses of interests are
\bq H_0: f(q) =0 \mbox{ for all } q \in   \mathcal{M}  \quad vs. \quad H_1: f(q) > 0 \mbox{ for some } q \in  \mathcal{M}. \label{eq:H0} \eq
Once incremental formula (\ref{eq:increment}) is identified, the $p$-value for pointwise inference at {\em each fixed} $q$ can be computed iteratively.  At the $k$-th random walk,  the $p$-value is given as $p_k$. Then $p_{k+1}$ is computed from iterative formula 
\bqn (k+1) p_{k+1} = kp_k  + 
 \mathcal{I} \Big( f ({\bf x},{\bf y}) \geq f(\pi_{ij}({\bf x}),\pi_{ij}({\bf y})) \Big),\label{eq:pk} \eqn
where $\pi_{ij}$ changes over random walks.
 Note the $p$-value for multiple comparisons over all $q$ is given by 
 \bq p\mbox{-value} =
P\Big(\bigcup_{q \in  \cal{M}} \{f(q) >
h\}\Big)= 
P \Big(\sup_{q \in
  \cal{M}} f(q) > h \Big)\eq
for some threshold $h$ \cite{worsley.1996}. Thus, $\sup_{q \in
  \cal{M}} f(q)$ is used as a test statistic for multiple comparisons. Then the pointwise iterative formula (\ref{eq:pk}) changes to
\bq (k+1) p_{k+1} = kp_k  + 
 \mathcal{I} \Big(  \sup_{q \in \mathcal{M}} f ({\bf x}(q), {\bf y}(q)) \geq  \sup_{q \in \mathcal{M}} f(\pi_{ij}({\bf x}(q)),\pi_{ij}({\bf y}(q))) \Big).  \label{eq:pk2} \eq
For alternate hypothesis $H_1: f(q) < 0$, a similar algorithm can be used for test statistic $\inf_{q \in \mathcal{M}} f (q)$ for multiple comparisons.

\section{Validation}

{\em Numerical accuracy.}  We simulated $x_1, \cdots, x_m  \sim 0.1 + Unif(0,1)$, $y_1, \cdots, y_n \sim Unif(0,1)$ independently for $m=n=40$, where $Unif(0,1)$ is a continuous uniform distribution between 0 and 1. We determined the numerical accuracy based on a half million sequential random walks. For two-sample $t$-statistic, at each random walk, we iteratively updated $t$-statistic using the proposed method. The final $t$-statistic,  at the end of half million walks, is compared against the ground truth, which is directly computed from the two sample $t$-statistic formula applied to data after half million walks. The absolute error is $( 4.15 \pm 4.29) \times 10^{-13}$ over 100 independent simulations. 
Similarly for twin-correlations, we also performed the above simulations 100 times and obtained the absolute error of $( 5.87 \pm 5.45) \times 10^{-13}$. The cumulating  error over a half million random walks is negligible.\\

\begin{figure}[t]
\centering
\includegraphics[width=1\linewidth]{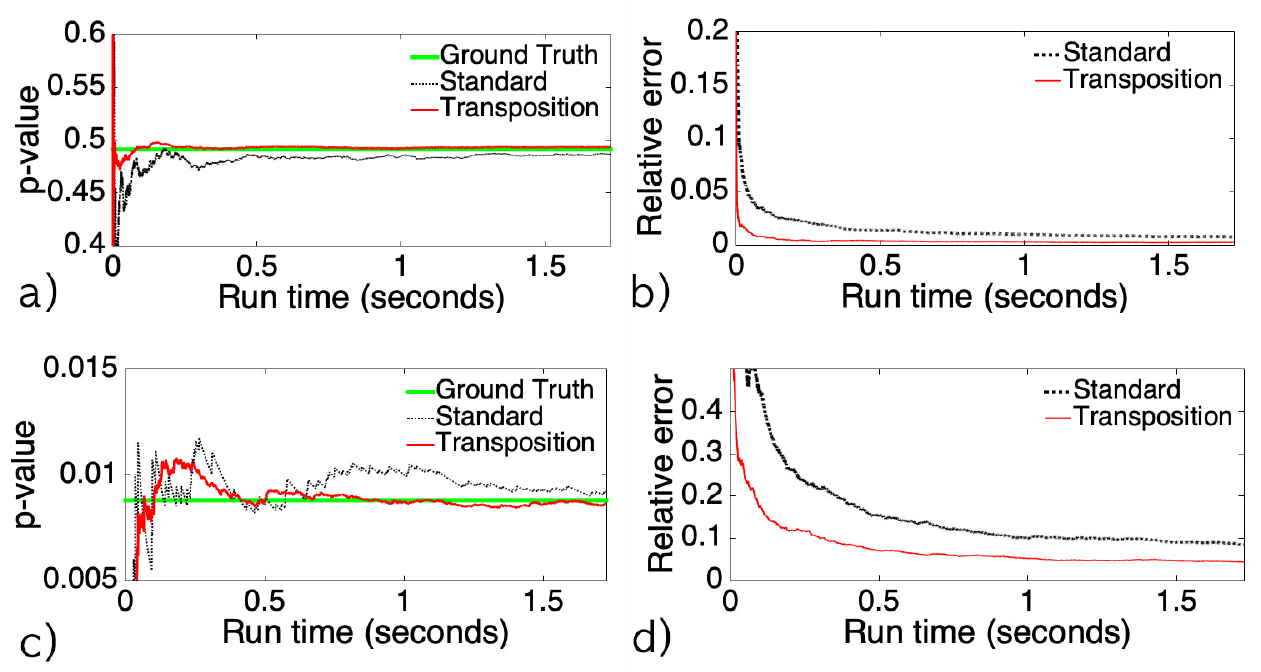}
\caption{\small a, c) One representative simulation study showing faster convergency of the transposition method. The Gaussian distribution provides the exact ground truth. b,d) The average relative error against the ground truth. The average of 100 independent simulations was plotted.}
\label{simulation}
\end{figure}

{\em Simulation setting.} We compared the proposed transposition method against the standard permutation test in random simulations with the ground truth. We simulated $x_1, \cdots,$ $x_{m} \sim N(0,1)$, standard normal distribution and $y_1, \cdots, y_{n} \sim 0.1+ N(0,1)$. Since the distribution is Gaussian, it provides the ground truth in computing $t$-statistic and $p$-value. The simulations were independently performed 100 times and their average was reported. \\

{\em Simulation 1} (small sample size). We used $m=n=10$. 
The sample sizes are too small to differentiate the group difference. We obtained the $t$-statistic value of -0.48, which corresponds to the exact $p$-value of 0.33  (Figure \ref{simulation}-a green line). We performed the standard permutation test with up to 10000 permutations,  which took 1.67 seconds per simulation in average. At the same run time, the transposition was sequentially done 1221900 times.  The transposition method converged faster than the standard permutation test through the whole run time due to 122 times more permutations the transposition method generated (Figure \ref{simulation}-b). The relative errors of the transposition method is about half of that of the standard method in most run time. \\

{\em Simulation 2} (large sample size). We used $m=n=100$. The sample sizes are big enough to differentiate the group difference. We obtained the $t$-statistic value of 2.39 and corresponding $p$-value of 0.0088, which are taken as the ground truth  (Figure \ref{simulation}-c green line). We performed the standard permutation test with up to 1 million permutations, which took 173 seconds per simulation in average.  With the same run time, the transposition was sequentially done about 125 million times.  The transposition method converged faster than the standard permutation test uniformly through the whole run time (Figure \ref{simulation}-d). The performance results did not change much even if we performed more permutations over longer durations with different simulation parameters.

The standard method is heavily influenced by the sample size while the rapid acceleration is not influenced by the sample size much due to constant run time in each walk.

\section{Application}
The proposed rapid acceleration of the permutation works for any type of permutation tests. Here we show two specific  applications examples as illustration. 

{\em Subjects.} We used the T1-weighted MRI of 228 twin pairs for the study from  the Human Connectome Project \cite{vanessen.2012}.  Excluding 11 twin pairs that are missing a subject, we used genetically confirmed 138 monozygotic (MZ) and 79 same-sex dizygotic (DZ) twin pairs. There are 274 females and 182 males in the dataset. The details on preprocessing can be found in \cite{smith.2013} that includes nonlinear image registration and white matter and pial surface mesh extractions in FreeSurfer.

\begin{figure}[t]
\centering
\includegraphics[width=1\linewidth,clip=true]{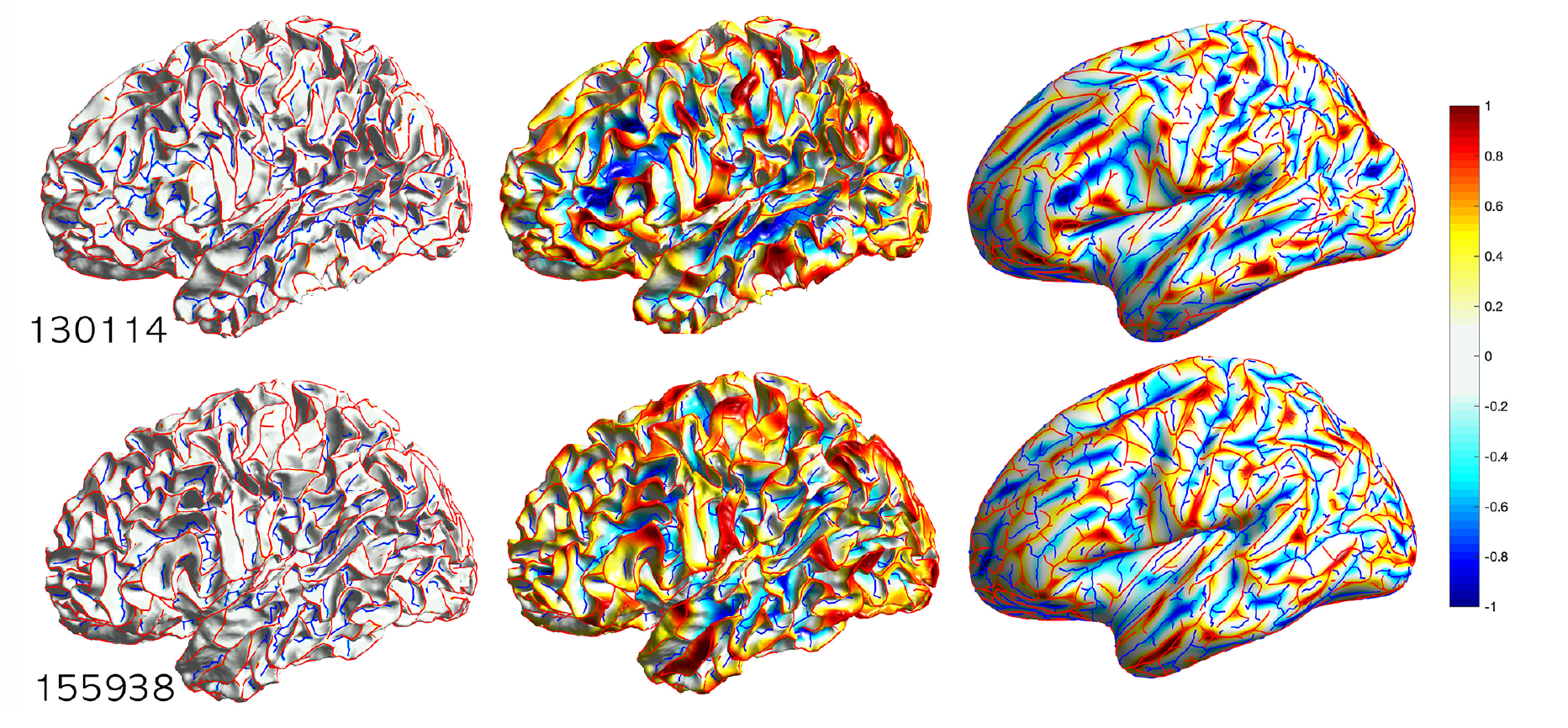}
\caption{ Sulcal (blue) and gyral (red) curves are extracted and displayed along the white matter surfaces for two subjects 130114 and 155938 (left). Gyral curves are assigned value 1 and sulcal curves are assigned value -1. All other parts of surface mesh vertices are assigned value 0. Then heat diffusion was performed with diffusion time 0.001 (middle). The diffusion maps are flattened to show the pattern of diffusion (right). The diffusion maps are used for statistical analysis in localizing the regions of the brain that differentiates male and female differences or identifying heritable brain regions.}
\label{fig:sulcalcurve}
\end{figure}

{\em Sulcal and gyral curves.} We use the automatic sulcal curve extraction method \cite{lyu.2018} that detects concave regions (sulcal fundi) along which sulcal curves are traced. The method follows two main steps: (1) sulcal point detection and (2) curve delineation by tracing the detected sulcal points. Sulcal points are determined by employing the line simplification method \cite{douglas.1973} that denoises the sulcal regions without significant loss of their morphological details. A partially connected graph is then constructed by the sulcal points, where edge weights are assigned based on geodesic distances. Finally, the sulcal curves are traced over the graph by the Dijkstra's algorithm \cite{dijkstra.1959}. Similarly, we extend this to gyral curve extraction by finding convex regions. 
The sulcal and gyral curves for two subjects 130114 and 155938 are shown on the white matter surfaces in Figure~\ref{fig:sulcalcurve}. 

{\em Diffusion maps.}  Gyral curves are assigned heat value 1 and sulcal curves are assigned heat value -1. All other parts of surface mesh vertices are assigned value 0. The difference in the initial temperature produces heat gradient. Then heat diffusion was performed with diffusion time 0.001 on brain surfaces by solving
\bq \frac{\partial f}{\partial t}(q,t)  = \Delta f(q,t), \label{eq:heat}\eq
where $\Delta$ is the Laplace-Beltrami operator on the surface. The solution to heat diffusion is numerically given by
$$f(q,t) = \sum_{i=0}^k e^{-\lambda_i t} f_i \psi_i(q), \; \quad
f_i =\int f(q,t) \psi_i(q)\; d\mu(q),$$
where $\lambda_j$ and $\psi_j$ are eigenvalues and eigenfunctions satisfying $\Delta \psi_i = -\lambda_i \psi_i$ \cite{chung.2007.TMI}. $f_i$ are Fourier coefficients with respect to orthonormal basis $\psi_i$. Enough terms, i.e., $k=10000$, are chosen to numerically guarantee the convergence. The diffusion map measures relative distance between sucal and gyral curves. If sulcal and gyral curves are in close proximity, heat will diffuse faster. 
The diffusion maps are subsequently used in localizing the regions of the brain that differentiate male and female differences or MZ- and DZ-twins.

\begin{figure}[t]
\centering
\includegraphics[width=1\linewidth,clip=true]{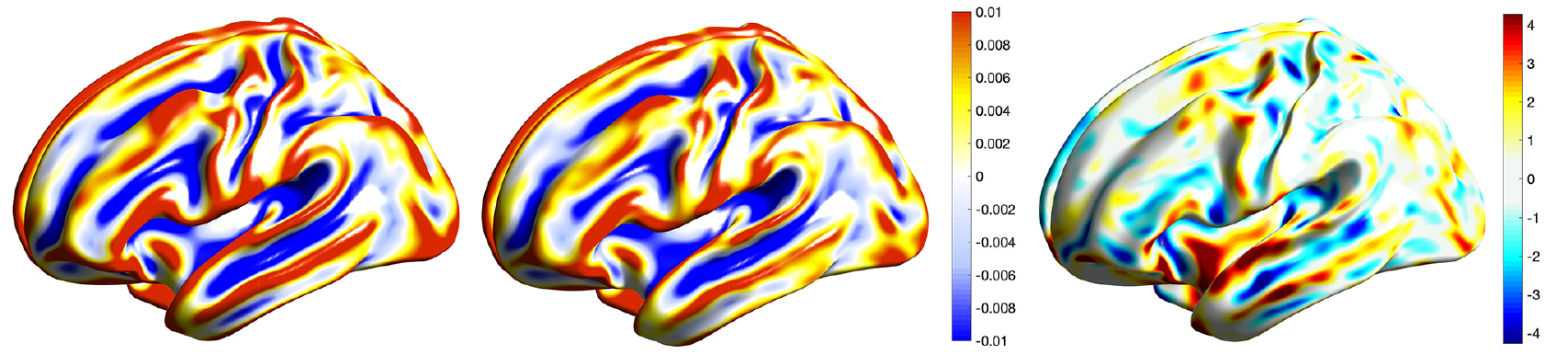}
\caption{The average diffusion maps of  274 females (left) and 182 males (middle) projected to the average surface template. The $t$-statistic map (right) shows localized sulcal and gyral pattern differences (female - male) thresholded at -4.27 and 4.48 (corrected $p$-values $<$ 0.05).
}
\label{fig:malefemale}
\end{figure}

{\em Statistical analysis (Female vs. Male).}  The data was split into females (274 subjects) and males (182 subjects). The average diffusion maps of females and males show major differences in the temporal lobe, which is responsible for processing sensory input into derived meanings for the appropriate retention of visual memory, language comprehension, and emotion association \cite{smith.2007}. The two-sample $t$-statistics were constructed on the diffusion maps. The maximum $t$-statistic map is 7.08 while the minimum $t$-statistic map is -6.44. The rapid acceleration method was used to generate half million permutations, which took  \underline{40 minutes} in a laptop computer. They were used to find the  distribution of maximum and minimum $t$-statistics over cortical surfaces for multiple comparisons. For $t$-statistic values larger than 4.48 and smaller than -4.27 gives $p$-value smaller than 0.05 (corrected). In comparison, it would taken about \underline{18 days} to generate half million permutations using the standard permutation method in the same computer.

\begin{figure}[t]
\centering
\includegraphics[width=1\linewidth,clip=true]{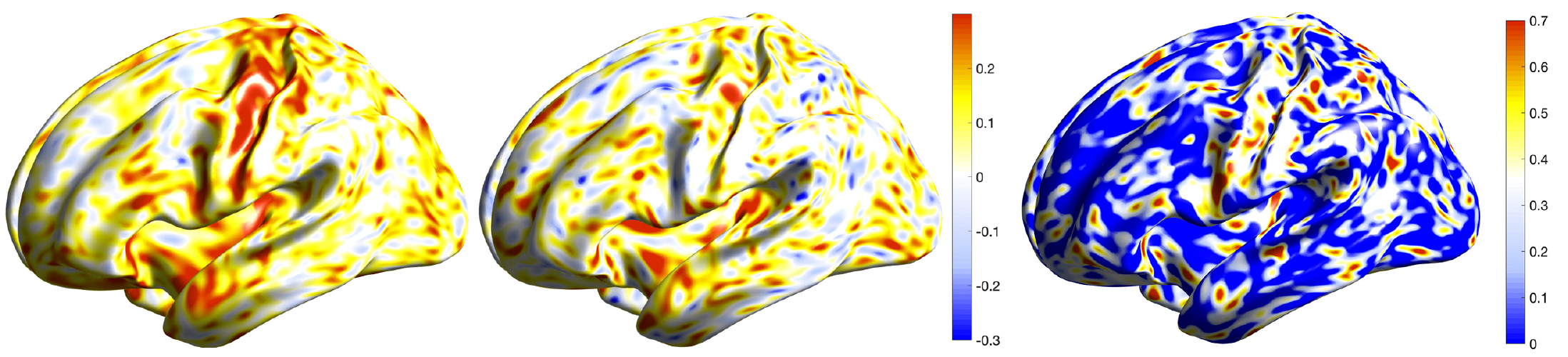}
\caption{
The averaged twin correlation maps of 138 MZ twin pairs (left) and  79 DZ twin pairs (middle) over 10000 random walks.
The heritability index (HI) map (right) shows the  amount of genetic contributions mainly concentrated on central gyral and sulcal areas.}
\label{fig:MZDZ}
\end{figure}

{\em Statistical analysis (MZ vs. DZ twins).}  The dataset also have twin information. There are 138 MZ twin pairs and 79 same-sex DZ twin pairs.
At each vertex position $p$, we computed the twin correlation of diffusion map, denoted by $C_p^{MZ}$ and $C_p^{DZ}$.
The heritability index (HI), the amount of genetic variations  due to genetic influence in a population, were was computed by $\textmd{HI}_p=C_p^{MZ}-C_p^{DZ}$ \cite{falconer.1995,chung.2017.IPMI}.
Since there is no ordering in twins, twins can be transposed
one pair at a time in random walks. Thus, we averaged 10000 twin correlations generated over random walks, which took \underline{97 seconds} for both twin groups to guarantee convergence up to 3 decimal places. Figure \ref{fig:MZDZ} displays the result of averaged correlation maps and the corresponding HI map. Although many regions are shown to be highly heritable (HI $> 0.7$), most heritable regions are localized in central gyral and sulcal areas.

\section{Discussion}
We have presented a new algorithm for the slow random walk based permutation test that is actually faster than the standard permutation test. 

We did not peruse every possible theoretical questions. One such question is {\em how many random walks are needed to reach all the permutations in  $\mathbb{S}_{n+m}$}? This is a nontrivial active research problem beyond the scope of this paper \cite{aldous.1983,bafna.1998}. 

Instead of performing random walks with the uniform sampling strategy, we can perform more structured random walks such that each walk produces distinct permutation from previous permutations. Such structured random walk will likely to speed up the convergence further. Although we did not show here, it is also possible to construct incremental procedure for computing other test statistics such as $F$-statistic and Spearman correlation over random walks.  These problems are left as future studies.

\section*{Acknowledgements}
This work was supported by NIH grant R01 EB022856. We thank Andrey Gritsenko of University of Wisconsin-Madison, Martin Styner of University of North Carolina and Li Shen of University of Pensilvania for valuable supports and discussions. 

\bibliographystyle{plain}
\bibliography{reference.2018.12.01}

\end{document}